\documentclass[journal]{IEEEtran}
\usepackage{subcaption}
\usepackage{kantlipsum} 
\usepackage{mwe} 
\usepackage{graphicx}
\usepackage{float}
\usepackage[T1]{fontenc}
\usepackage{dcolumn}
\usepackage{color}

\usepackage[hyphens]{url}

\usepackage[noadjust]{cite}
\usepackage[cmex10]{amsmath}

\newcolumntype{d}[1]{D{.}{.}{#1}}
\DeclareCaptionLabelSeparator{periodspace}{.\quad}
\begin{document}
\title{Back-of-the-envelope evaluation of the prevalence of \\RIMP~or~LOS~propagation~as~a~function~of~frequency}

\author{\IEEEauthorblockN{
	Aidin~Razavi,
	Andr\'{e}s~Alay\'{o}n~Glazunov,~\IEEEmembership{Senior Member,~IEEE},
	Rob~Maaskant,~\IEEEmembership{Senior Member,~IEEE},
	and Jian~Yang,~\IEEEmembership{Senior Member,~IEEE}
}
\thanks{This work has been supported by two projects from Sweden's innovation agency VINNOVA, one within the VINN Excellence Center Chase at Chalmers and another via the program Innovative ICT 2013, and by internal support from Chalmers.

A.~Razavi was with the Signals and Systems Department, Chalmers University of Technology at the time this work was conducted. He is now with Ericsson AB, Sweden. A.~A.~Glazunov is with the Department of Electrical Engineering, University of Twente, P.O. Box 217, 7500 AE Enschede, The Netherlands, R.~Maaskant and J.~Yang are with the Electrical Engineering Department, Chalmers University of Technology, SE-41296 Gothenburg, Sweden (e-mail: aidin.razavi@tgeik.com; a.alayonglazunov@utwente.nl; rob.maaskant@chalmers.se;~jian.yang@chalmers.se).~A.~A.~Glazunov is also affiliated with the Department of Electrical Engineering, Chalmers University of Technology, (andres.glazunov@chalmers.se), R.~Maaskant is also with the Department of Electrical Engineering, Eindhoven University of Technology (TU/e), (r.maaskant@tue.nl).}}
\maketitle
\label{Paper:Generic}
\begin{abstract}
The performance of 5G wireless communication systems, employing Massive-MIMO at millimeter-wave frequencies, is most likely measured only in Over-The-Air (OTA) setups.
It is proposed to perform OTA measurements in two limiting environments of Rich Isotropic MultiPath (RIMP) and Random Line-of-Sight (Random-LOS) instead of a typical or representative channel. In the present paper, we present a back-of-the-envelope investigation of the impact of scattering on the frequency dependence of the signal fading statistics in the $500$~MHz--$100$~GHz band. We introduce a simple model for a generic scattering environment by using randomly distributed resonant scatterers to investigate the impact of the size of the scattering environment, the scatterer density, and the number of scatterers on the signal variability in terms of the Rician $K$-factor as a function of frequency. The simplified model is also verified against full-wave simulation using the Method of Moments (MoM).
\end{abstract}

\begin{IEEEkeywords}
	Propagation, Scattering, Scattering parameters, Antenna measurements, Fading channels.
\end{IEEEkeywords}

\section{Introduction}
$5$G wireless communication networks are being developed to meet the increasing demands for better quality-of-service, e.g., throughput in the Gbps range. The multi-user multiple-input multiple-output (MU-MIMO) technology employing very large array antennas at the radio base station (RBS), also known as Massive MIMO technology, is one of the key technology enablers \cite{boccardi2014five}. With this technology, the RBS will serve many mobile stations simultaneously, and the data reaches each of the mobile antennas by beamforming the energy towards them \cite{larsson2014massive}. The combination of large array antennas and many users turns out to be favorable for data transmission. Favorable propagation (FP) conditions means that the channel vectors between the users and the RBS are nearly pairwise orthogonal. Therefore, the signal processing complexity can be considerably reduced since linear processing is very close to {\color{black} be} optimal \cite{marzetta2010noncooperative,rusek2013scaling}.

Another enabler is the use of very large portions of the electromagnetic spectrum. That is, in addition to operation in the traditional ultra high frequency (UHF) band, new very large contiguous bandwidths will be exploited at the extremely high frequency (EHF) band, i.e., the millimeter-wave (mm-wave) frequencies.

Currently, there is no complete understanding of the propagation characteristics at the frequency bands of interest, i.e., from $500$~MHz to approximately $100$~GHz for Massive MIMO. As is well-known, channel models are indispensable tools for evaluating, predicting or optimizing the performance of wireless systems \cite{glazunov2012mimo}. However, {\color{black} it is in general not fully possible to know exactly what channel model is the typical, or the most representative one.} On the other hand, it is often useful to look at phenomena in the limiting cases to model certain phenomena, e.g., static or high frequency limits.

In wireless communications, the {\color{black}Rich Isotropic MultiPath} (RIMP) propagation channel and the Line-of-Sight (LOS) propagation channel represent two limiting propagation environments in terms of the spatial distribution of the angle-of-arrival (AoA) or angle-of-departure (AoD). As argued in \cite{ngo2014aspects, bjornson2015massive}, both channels are favorable for the operation of Massive MIMO. Therefore, it is reasonable to expect that real propagation environments, which are likely to be in-between these extremes, would also be favorable. This has support from experimentally observed FP characteristics of Massive MIMO channels in real life \cite{gao2015massive}.

This fact has a very profound implication, i.e., both the RIMP and the LOS environments may suffice to characterize the OTA performance of wireless devices as suggested in \cite{kildal2013rethinking}. The following \emph{real-life hypothesis} for OTA device characterization has been formulated {\color{black}as}: {\em if a wireless device is proven to work well in RIMP and Random-LOS, it will work well in all real-life environments}~\cite{kildal2013new}. It is worthwhile to note that the Random-LOS propagation environment is a generalization of the LOS environment concept, where the randomness is a result of the unpredictable positions and orientations of antennas of the mobile stations, the deployment position of a RBS, or both. In Random-LOS, both the Angle-of-Arrival (AoA) and the polarization of the LOS wave (i.e., the only wave present) are considered to be random variables.

The RIMP and the Random-LOS environments are especially relevant to the OTA characterization of Massive MIMO RBS in 5G wireless systems. 5G massive array antennas will, in practice, be only possible to measure in OTA setups due to the large number of ports. Moreover, as we go higher in frequency, e.g., for systems operating at mm-wave frequencies, there will be most likely no access to measurement ports at all.

Here we therefore present a ``back-of-the-envelope" investigation of the impact of scattering on the signal fading statistics as a function of frequency from $500$ MHz to $100$ GHz. We neglect the water vapour and oxygen absorption effects as well as the propagation mechanisms leading to large-scale signal fading fluctuations. We assume that these will occur on top of the presented scattering model. We develop a model for signal fluctuations due to scattering, i.e., short-term fading, as a function of frequency under various simplifying assumptions. The main idea is to investigate the impact of the size of the scattering environment, the scatterer density, the number of scatterers on the signal fading (i.e., signal variability) as a function of frequency. The model is general{\color{black} and is not} derived for specific type of scatterers. Modeling the scatterers {\color{black}by} resonant dipoles, the analytical {\color{black}model} is compared to numerical computations performed with functions inherited from the CAESAR code~\cite{maaskant2010analysis}. Under {\color{black}the assumptions used in this work}, it is shown that as we go higher in frequency, the power in the LOS component gradually increases as compared to the power of the scattering contribution. {\color{black}This phenomenon has recently been observed in measurements as well~\cite{5g-mmmagic}.} Hence, for a fixed number of thin wire scatterers, the scattering environment behaves like the RIMP channel at lower frequencies, while at higher frequencies it becomes more like the (Random-)LOS channel.

{\color{black}The paper is organized as follows. In Sect. II, the signal fading model and Rician $K$-factor are defined. Then, the scattering model and all assumptions for this model are presented in Sect. III. The detailed derivation of $K$-factor as a function of the average scattering cross-section of scatterers and therefore its dependence on the frequency are provided in Sect. IV. The results from the analytical $K$-factor computations and a MoM numerical simulations for different cases are compared in Sect. V, with discussions and analysis on the results. Finally, the paper is concluded in Sect. VI.}

\section{Signal Fading Model}
\begin{figure}[!t]
	\centering
	\includegraphics[width=0.8\columnwidth]{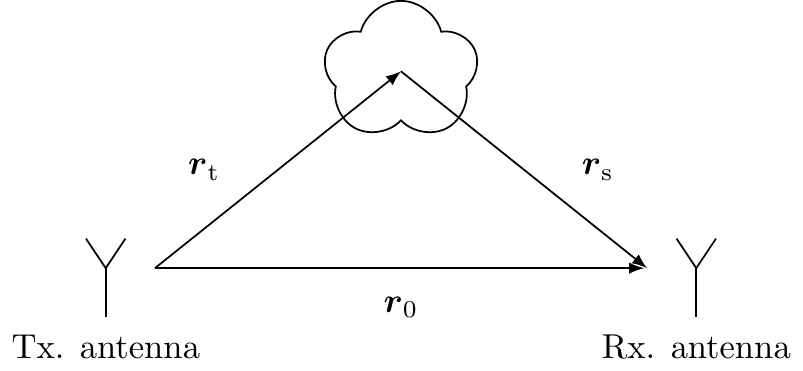}
	\caption{\small Transmitter and receiver antennas in the scatterering environment.}
	\label{fig:model}
\end{figure}

To study the scattering environment, let us assume a transmitter and a receiver antenna in the presence of scatterers as illustrated in Fig.~\ref{fig:model}. In order to model the single-port receive signal we introduce the complex random variable
\begin{align}
v=\frac{V_{\mathrm{oc}}}{2\sqrt{2R_{\mathrm{ar}}}},\label{eq:3}
\end{align}
where $V_{\mathrm{oc}}$ is the \emph{total} open-circuit voltage induced at the receive antenna ports and $R_{\mathrm{ar}}$ is the real part of the receive antenna input impedance.
The received power as a function of the open circuit voltage for the conjugate-matched load condition is
\begin{align}\label{eq:4}
P_\mathrm{ar}=|v|^2=\frac{|V_{\mathrm{oc}}|^2}{8R_{\mathrm{ar}}}.
\end{align}

Scattering, {\color{black}including physical phenomena such as reflection, refraction and diffraction,} is the dominating propagation mechanism in wireless multipath contributions \cite{glazunov2012mimo}. Objects (scatterers) surrounding the mobile antenna are assumed to contribute the most to the fast fading fluctuations of the received signal \cite{jakes1974microwave}. Hence, we seek to estimate the impact of the scattering on the fast fading process as a function of frequency.

The Rician probability distribution function (pdf) has gained widespread acceptance as a model of the continuous wave signal fluctuations caused by multipath propagation. A measure of the severity of fluctuations is then given by the Rician $K-$factor defined as
\begin{equation}
K=\frac{P_\mathrm{LOS}}{P_\mathrm{RIMP}}=\frac{|\langle v \rangle|^2}{\langle|v|^2\rangle-|\langle v\rangle|^2},\label{eq:1}
\end{equation}
where $P_\mathrm{LOS}$ and $P_\mathrm{RIMP}$ are the powers of the LOS and the RIMP components, respectively, $\langle v\rangle$ denotes the ensemble average of complex-valued random variable $v$.

The second parameter that defines the Rician pdf is the total received power
\begin{equation}
P_\mathrm{r}=P_\mathrm{LOS}+P_\mathrm{RIMP}=\langle|v|^2\rangle.\label{eq:2}
\end{equation}

The Rician distribution of the envelope of the complex signal amplitude received by the antenna $|v|$ is given by \cite{jakes1974microwave},
\begin{eqnarray}\label{eq:5}
f_{|v|}(|v|)&=&\frac{2\left(1+K\right)|v|}{P_{\mathrm{r}}}\exp\left(-K-\frac{(K+1)|v|^{2}}{P_{\mathrm{r}}}\right)\nonumber \times \\ &&I_{0}\left(2\sqrt{\frac{K(1+K)}{P_{\mathrm{r}}}}|v|\right),
\end{eqnarray}
where $I_{0}$ is the modified Bessel function of the first kind and zeroth order.

The physical interpretation of \eqref{eq:1} is that the receive signal fading, or severity of fluctuations, is a function of the proportion between the power of the deterministic component of the received signal, i.e., given by the numerator and the power in the stochastic component, i.e., given by the denominator. The former can be interpreted as the LOS field component while the latter can be interpreted as the RIMP component. Two limiting cases immediately arise, i.e., $K\rightarrow 0$ and $K\rightarrow\infty$ denoting the RIMP and the LOS channels, respectively. Thus, intermediate values of $K$ will describe Rician propagation channels in between the RIMP and the LOS channels. It is worthwhile to note that in practice, the deterministic component can be just a strong reflected or diffracted wave, while the stochastic component not necessarily arises from an isotropic wave field distribution. For the sake of simplicity we will not consider this general case. Instead, we will look into the situation when there is a LOS field component in addition to a RIMP field component.

\section{Scattering Model and General Assumptions}
In wireless communication channels, as well as in microwave sensing or radar applications, there is usually more than one single scatterer interacting with the receive and transmit antennas. Many scatterers have to be considered in order to completely define the propagation channel. However, in practice, full knowledge about the exact physical
properties and positions of the scatterers is not available. Here, we are mainly concerned with the statistical modeling of the channel; an approach that has gained widespread use in wireless communication \cite{molisch2012wireless} and remote sensing applications \cite{tsang2004scattering}. For this reason, the antennas and scatterers are assumed to be in the far-field region of each other and single scattering is assumed.

To fully and exactly evaluate the scattering contribution to the total received signal is a too complex task to accomplish even under the assumptions already stated above. We need therefore to introduce further assumptions. The total open-circuit voltage entering \eqref{eq:3} is a random variable due to the random nature of the scattered field component as shown further below. The randomness arises from the random positions of the scatterers relative to each other and the antennas. In turn, this results in the polarization of the scattered waves {\color{black}behaving} like a random process too. Also the phase difference can be considered random due to different path lengths traveled by the scattered waves.

In a real scattering environment, the different scattering objects will appear to have different sizes depending on the frequency. Also the electrical distance between scatterers will increase with frequency. Moreover, different parts of a larger scatterer may be modeled by a set of smaller scatterers with no electrical coupling to each other under the approximations assumed herein. Small half-wavelength dipole antennas have been used to model the electromagnetic scattering in wireless channels in \cite{bialkowski2006investigating}. Herein, we adopt a similar approach.
Before we proceed further, let us summarize our simplifying modeling assumptions:

\begin{description}
  \item[(I)] We consider a narrowband continuous wave signal.
  \item[(II)] Antennas and scatterers are in the far-field of each other where the far-field reference distance is given by $R_\mathrm{FF}$ computed according to the criterion given in \cite{kim2013generalised}.
  \item[(III)] For the sake of simplicity, we assume that the scatterers are uniformly distributed within a spherical scattering volume. The probability distribution of a scatterer being located $\rho$ from the center of a sphere of radius $R_\mathrm{s}$ or at the angle $\theta$ is given by
\begin{subequations}
\begin{align}
p(\rho)=\frac{3\rho^2}{R_\mathrm{s}^3},\\
p(\theta)=\frac{\sin\theta}{2},
\label{eq:6}
\end{align}
\end{subequations}
respectively, where the radius $R_\mathrm{s}$ delimits the volume containing the scatterers. Furthermore, the receiver antenna is located at the center of the volume.
  \item[(IV)] We assume that the transmit antenna is far away from the scattering volume, while the receive antenna is within the scattering volume at $\rho=0$. {\color{black} The communication scenario assumes a local cloud of scatterers surrounding the mobile user which is distant to the base station antenna and therefore illuminated uniformly by a plane wave.}
  \item[(V)] Only single scattering is considered in {\color{black}our analytical model}. However, in the later MoM simulations multiple scattering is included.
  \item[(VI)] The scattering contribution of the scatterers is determined by their spatially averaged scattering cross-section.
  \item[(VII)] The randomness in the positions of scatterers is assumed to originate from an ensemble of states originating from a random process.
\end{description}

{\color{black} It is worthwhile to note that because our assumed channel model is simple in nature it can only predict dominant effects and thus general qualitative trends, such as the indication that the Rician K-factor increases with frequency, which is a phenomenon that has recently been observed in measurements as well~\cite{5g-mmmagic}. However, to predict more quantitatively what happens in other communication scenarios as well as in more specific environments, we refer e.g. to~\cite{Riaz2015,Ghoraishi2006,Samimi2016}.}

\section{Computation of $K-$factor and $P_\mathrm{r}$}
Consider the following scattering problem where the field radiated by a transmit antenna is scattered by a single linear scatterer. We are then interested in computing the total field propagating towards the receive antenna. The geometry of the problem is illustrated in Fig.~\ref{fig:model}.
The field radiated by the transmit antenna $\boldsymbol{E}_{\mathrm{d}}( \boldsymbol{r}_{\mathrm{t}}) $ is defined in the
far-field region by the far-field function $\boldsymbol{G}_{\mathrm{t}}(\boldsymbol{\hat{r}}_{\mathrm{t}}) $ radiated in direction $\boldsymbol{\hat{r}}_{\mathrm{t}}=\boldsymbol{r}_{\mathrm{t}}/r_{\mathrm{t}}$ {\color{black}as}
\begin{equation}\label{eq:8}
\boldsymbol{E}_{\mathrm{d}}( \boldsymbol{r}_{\mathrm{t}}) =
\boldsymbol{G}_{\mathrm{t}}( \boldsymbol{\hat{r}}_{\mathrm{t}}) \frac{
\mathrm{e}^{-jkr_{\mathrm{t}}}}{r_{\mathrm{t}}}+\mathcal{O}( r_{
\mathrm{t}}^{-2}) \text{ as }r_{\mathrm{t}}\rightarrow \infty,
\end{equation}
where $\mathcal{O}(x^n)$ stands for ``order of" asymptotic. This is the field impinging on the scatterer.

Similarly, in the far-field region, the scattered electric field \ $\boldsymbol{E}_{\mathrm{s}}$ is fully described by the far-field function {\color{black}$\boldsymbol{G}_{
\mathrm{s}}$} scattered in direction $\boldsymbol{\hat{r}}_{\mathrm{s}}=\boldsymbol{r}_{\mathrm{s}}/r_{\mathrm{s}}$ as
\begin{equation}\label{eq:9}
\boldsymbol{E}_{\mathrm{s}}(\boldsymbol{r}_{\mathrm{s}})=\boldsymbol{G}_{\mathrm{s}}( \boldsymbol{\hat{r}}_{\mathrm{s}}) \frac{\mathrm{e}^{-jkr_{\mathrm{s}}}}{
r_{\mathrm{s}}}+\mathcal{O}( r_{\mathrm{s}}^{-2}) \text{ as }r_{
\mathrm{s}}\rightarrow \infty,
\end{equation}
where $\boldsymbol{G}_{\mathrm{s}}$ can be expressed in terms of the scattering matrix $\mathcal{S}( \boldsymbol{\hat{r}}_{\mathrm{s}},
\boldsymbol{\hat{r}}_{\mathrm{t}})$ \cite{van2007electromagnetic}
\begin{equation}\label{eq:10}
\boldsymbol{G}_{\mathrm{s}}( \boldsymbol{\hat{r}}_{\mathrm{s}}) =
\mathcal{S}( \boldsymbol{\hat{r}}_{\mathrm{s}},\boldsymbol{\hat{r}}_{
\mathrm{t}}) \cdot \boldsymbol{E}_{\mathrm{d}}( \boldsymbol{r}_{
\mathrm{t}}),
\end{equation}
where we have assumed that the amplitude of the plane wave incident at the scatterer from direction $\boldsymbol{\hat{r}}_{\mathrm{t}}$ is given by $
\boldsymbol{E}_{\mathrm{d}}( \boldsymbol{r}_{\mathrm{t}}) $. Hence,
from \eqref{eq:8}-\eqref{eq:10}, the scattered field can be expressed as
\begin{equation}\label{eq:11}
\boldsymbol{E}_{\mathrm{s}}( \boldsymbol{r}_{\mathrm{s}}) = \mathcal{S}( \boldsymbol{\hat{r}}_{\mathrm{s}},\boldsymbol{\hat{r}}_{\mathrm{t}}) \cdot \boldsymbol{G}_{\mathrm{t}}
( \boldsymbol{\hat{r}}_{\mathrm{t}}) \frac{\mathrm{e}^{
-jk( r_{\mathrm{s}}+r_{\mathrm{t}}) }}{r_{\mathrm{s}}r_{
\mathrm{t}}}\text{ as }r_{\mathrm{s}},r_{\mathrm{t}}\rightarrow \infty
\end{equation}

The total field impinging on the receive antenna will then be the sum of the scattered field \eqref{eq:11} and the field radiated by the antenna \eqref{eq:8} in the direction of vector $\boldsymbol{r}_{\mathrm{o}}$
\begin{equation}\label{eq:12}
\boldsymbol{E}_{\mathrm{tot}} = \boldsymbol{E}_{\mathrm{d}}(\boldsymbol{r}_{\mathrm{o}})+\boldsymbol{E}_{\mathrm{s}}(\boldsymbol{r}_{\mathrm{s}}).
\end{equation}

The open-circuit voltage induced at the receive antenna ports by an impinging wave is given by \cite{kildal2015foundations}
\begin{equation}
V_{\mathrm{oc}}^{\mathrm{tot}}=\frac{2\lambda}{j\eta I} \boldsymbol{G}_{\mathrm{r}} \!\cdot\! \boldsymbol{E}_{\mathrm{tot}}, \label{eq:13}
\end{equation}
where $\boldsymbol{G}_{\mathrm{r}}$ is the far-field function of the receive antenna in
the direction $\boldsymbol{\hat{r}}$ for excitation current $I$ {\color{black}and} $\boldsymbol{E}_{\mathrm{tot}}$ the amplitude of the incident plane wave field measured at the phase center of the receive antenna that we choose to be the origin of the coordinate system associated with that antenna.

From \eqref{eq:12} and \eqref{eq:13} we see that the total open-circuit voltage can be expressed as the sum of two terms
\begin{equation}
V_{\mathrm{oc}}^{\mathrm{tot}}=V_{\mathrm{oc}}^{\mathrm{d}}+V_{\mathrm{oc}}^{\mathrm{s}}, \label{eq:14}
\end{equation}
where both terms are obtained from the corresponding terms in \eqref{eq:12}, i.e., induced by the direct field and the scattered field, respectively.

We consider next the situation when many scatterers are present between the transmit and the receive antennas. The scatterers and the antennas have associated with them local coordinate systems, while their relative positions are defined relative to a common coordinate system. Single scattering (no coupling between the scatterers) is assumed, unlike in the later MoM simulations. Enforcing the above made assumptions gives after some algebraic manipulations the following expression for the open-circuit voltage induced by $N_\mathrm{s}$ scatterers
\begin{align}\label{eq:15}
V_{\mathrm{oc}}^{\mathrm{s}}\!&=\!\frac{2\lambda}{j\eta I}\!\sum_{n=1}^{N_\mathrm{s}} \boldsymbol{G}_{\mathrm{r}}( \boldsymbol{\hat{r}}_{\mathrm{r}n})\!\cdot\!\mathcal{S}_n( -\boldsymbol{\hat{r}}_{
\mathrm{r}n},\boldsymbol{\hat{r}}_{\mathrm{t}n})\!\cdot\!\boldsymbol{G}_{
\mathrm{t}}( \boldsymbol{\hat{r}}_{\mathrm{t}n})
\frac{\mathrm{e}^{-jk( r_{\mathrm{r}n}+r_{\mathrm{t}n}) }}{r_{\mathrm{r}n}r_{\mathrm{t}n}},
\end{align}
where we use the radius-vector notation introduced in Fig.~\ref{fig:model}. We have also introduced the sub-index $n$ in the scattering matrix to indicate that scatterers are in general different. The open-circuit voltage for the direct wave is given by
\begin{align}\label{eq:16}
V_{\mathrm{oc}}^{\mathrm{d}}\!&=\!\frac{2\lambda}{j\eta I}\boldsymbol{G}_{\mathrm{r}}( -\boldsymbol{\hat{r}}_{\mathrm{o}})\!\cdot\!\boldsymbol{G}_{
\mathrm{t}}( \boldsymbol{\hat{r}}_{\mathrm{o}})\frac{\mathrm{e}^{-jkr_{\mathrm{o}}}}{r_{\mathrm{o}}}.
\end{align}
Under the assumptions stated above, \eqref{eq:14}-\eqref{eq:16} provide a rather general description of the signal scattering model satisfying the above-stated assumptions.

In order to evaluate \eqref{eq:1}, we need to evaluate $\langle V_{\mathrm{oc}}^{\mathrm{tot}}\rangle$ and $\langle|V_{\mathrm{oc}}^{\mathrm{tot}}|^2\rangle$ first. As can be seen from \eqref{eq:15}, $\langle V_{\mathrm{oc}}^{\mathrm{tot}}\rangle=\left< V_{\mathrm{oc}}^{\mathrm{s}}\right>+V_{\mathrm{oc}}^{\mathrm{d}}$. Hence, we need to find
\begin{align}\label{eq:17}
\langle V_{\mathrm{oc}}^{\mathrm{s}}\rangle\!&=\!\sum_{n=1}^{N_\mathrm{s}} \Big\langle \frac{2\lambda}{j\eta I} \boldsymbol{G}_{\mathrm{r}}\!\cdot\!\mathcal{S}_n\!\cdot\!\boldsymbol{G}_{
\mathrm{t}}\Big\rangle \Big\langle\frac{\mathrm{e}^{-jk( r_{\mathrm{r}n}+r_{\mathrm{t}n}) }}{r_{\mathrm{r}n}r_{\mathrm{t}n}}\Big\rangle,
\end{align}
where we have omitted the arguments of some functions for the sake of simplicity. Observe that the ensemble averaging has been factored into two terms: (i) a term that comprises the scattering matrix, which describes randomness of the scattered field polarization and AoA at the location of the receiver antenna and, (ii) a term comprising the random positions of the scatterers. Both are independent random processes. We immediately see that
\begin{align}\label{eq:18}
\langle V_{\mathrm{oc}}^{\mathrm{s}}\rangle\!&=0,
\end{align}
since both ensemble averages are zero as shown in Appendix A. Hence, we obtain that
\begin{equation}\label{eq:19}
\langle V_{\mathrm{oc}}^{\mathrm{tot}}\rangle=V_{\mathrm{oc}}^{\mathrm{d}}
\end{equation}
Then from \eqref{eq:4}, \eqref{eq:16} and the Friis equation \cite{kildal2015foundations} we arrive at
\begin{align}\label{eq:20}
P_\mathrm{LOS}=\frac{|\langle V_{\mathrm{oc}}^{\mathrm{tot}}\rangle|^2}{8R_{\mathrm{ar}}}=\frac{|V_{\mathrm{oc}}^{\mathrm{d}}|^2}{8R_{\mathrm{ar}}}= \left(\frac{\lambda}{4 \pi r_\mathrm{o}}\right)^2 G_{\mathrm{or}}G_{\mathrm{ot}}P_\mathrm{t},
\end{align}
where $G_{\mathrm{or}}$, $G_{\mathrm{ot}}$ are the gains of the receive and transmit antennas in the LOS direction and $P_\mathrm{t}$ is the transmit power.

We see from \eqref{eq:14} and \eqref{eq:18} that
\begin{equation}\label{eq:21}
\langle|V_{\mathrm{oc}}^{\mathrm{tot}}|^2\rangle=\langle|V_{\mathrm{oc}}^{\mathrm{s}}|^2\rangle+|V_{\mathrm{oc}}^{\mathrm{d}}|^2.
\end{equation}
Hence, we need to find
\begin{align}\label{eq:22}
\langle|V_{\mathrm{oc}}^{\mathrm{s}}|^2\rangle\!&=\!\sum_{n=1}^{N_\mathrm{s}} \sum_{n^\prime=1}^{N_\mathrm{s}}\Big\langle \left(\!\frac{2\lambda}{\eta |I|}\!\right)^2 \left(\boldsymbol{G}_{\mathrm{r}}\!\cdot\!\mathcal{S}_n\!\cdot\!\boldsymbol{G}_{
\mathrm{t}}\right)\left(\boldsymbol{G}_{\mathrm{r}}\!\cdot\!\mathcal{S}_{n^\prime}\!\cdot\!\boldsymbol{G}_{
\mathrm{t}}\right)^\ast\Big\rangle \nonumber \times \\
& \Big\langle\frac{\mathrm{e}^{-jk( r_{\mathrm{r}n}+r_{\mathrm{t}n}-r_{\mathrm{r}n^{\prime}}-r_{\mathrm{t}n^{\prime}}) }}{r_{\mathrm{r}n}r_{\mathrm{t}n}r_{\mathrm{r}n^{\prime}}r_{\mathrm{t}n^{\prime}}}\Big\rangle,
\end{align}
After taking into account the random position of scatterers and the antenna parameters conventions and definitions in \cite{kildal2015foundations} we show in Appendix B that
\begin{align}\label{eq:23}
P_\mathrm{RIMP}=\frac{\langle|V_{\mathrm{oc}}^{\mathrm{s}}|^2\rangle}{8R_{\mathrm{ar}}}\!&=\!  \frac{3N_\mathrm{s}}{4\pi R_\mathrm{s}^2}\left(\frac{\lambda}{4 \pi r_\mathrm{o}}\right)^2 \frac{e_{\mathrm{r}}}{2}G_{\mathrm{ot}}\langle \sigma_\mathrm{s}\rangle P_\mathrm{t},
\end{align}
where $e_{\mathrm{r}}$ is the radiation efficiency of the receive antenna; all the other variables have been defined above.

Combining \eqref{eq:20} and \eqref{eq:23} into \eqref{eq:1} we obtain an estimate of the frequency dependence of the $K-$factor as function of frequency
\begin{align}\label{eq:24}
K(f)=\frac{8\pi D_{\mathrm{or}}R_\mathrm{s}^2}{3N_\mathrm{s}\langle \sigma_\mathrm{s}\rangle},
\end{align}
where we have used the relationship between antenna gain and directivity $G_{\mathrm{or}}=e_{\mathrm{r}}D_{\mathrm{or}}$. Under the above assumptions, the frequency dependence of the $K-$factor is completely determined by the directivity of the receiver antenna and the type of scatterer used in the model, i.e., the corresponding $\langle \sigma_\mathrm{s}\rangle$. Expression \eqref{eq:24} describes the dependence of the $K-$factor on the radius of the scattering volume $R_\mathrm{s}$ for constant $N_\mathrm{s}$. Clearly, in this case the scatterers will be further away from the receiving antenna if $R_\mathrm{s}$ increases. This leads to a weaker contribution to the total scattered field power and therefore a predominance of the LOS component over the RIMP component, i.e., a larger $K-$factor.

Let us now keep the scatterer density $\rho_\mathrm{s}=N_\mathrm{s}/V_\mathrm{s}$ constant; where $V_\mathrm{s}=4 \pi R_\mathrm{s}^3/3$. Then, the $K-$factor can be written as
\begin{align}\label{eq:25}
K(f)=\frac{2D_{\mathrm{or}}}{\rho_\mathrm{s}R_\mathrm{s}\langle \sigma_\mathrm{s}\rangle}.
\end{align}

In this case, the trend is the opposite. Indeed, keeping the density constant, a larger radius of the scattering volume will result in a smaller $K-$factor due to the larger contribution of the RIMP component to the total received power as compared to the LOS component.

Results \eqref{eq:24} and \eqref{eq:25} are both obtained under the assumption that all scatterers and the antennas are in the far-field of each other. Now, for a fixed $R_\mathrm{s}$ there will be a maximum number of scatterers $N_\mathrm{s}$ that can be ``packed'' into this volume. To obtain this estimate we use the far-field distance $R_\mathrm{FF}$ to model the diameter of an imaginary sphere surrounding the scatterer. Hence, we need to estimate the number of spheres with volume $V_\mathrm{FF}=\pi R_\mathrm{FF}^3/6$ that can be packed into the volume containing the scatterers $V_\mathrm{s}=4 \pi R_\mathrm{s}^3/3$. This number is given by
\begin{align}\label{eq:26}
N_\mathrm{s}&=\frac{\eta_\mathrm{pack}V_\mathrm{s}}{V_\mathrm{FF}}=8\eta_\mathrm{pack}\left(\frac{R_\mathrm{s}}{R_\mathrm{FF}}\right)^3,
\end{align}
where $\eta_\mathrm{pack}\approx0.64$ is the packing density of random close packing of spheres \cite{jaeger1992physics}. The corresponding scatterer density becomes
\begin{align}\label{eq:27}
\rho_\mathrm{s}&=\frac{N_\mathrm{s}}{V_\mathrm{s}}=\frac{6\eta_\mathrm{pack}}{\pi R_{\mathrm{FF}}^3},
\end{align}

Thus, \eqref{eq:24} and \eqref{eq:25} both reduce to
\begin{align}\label{eq:28}
K(f)=\frac{\pi D_{\mathrm{or}}R_\mathrm{FF}^3}{3\eta_\mathrm{pack}R_\mathrm{s}\langle \sigma_\mathrm{s}\rangle},
\end{align}
which provides a lower bound on the $K-$factor in the scattering propagation environment described above. It is worthwhile to note that the far-field distance $R_\mathrm{FF}$ also depends on the frequency \cite{kim2013generalised}
\begin{equation}\label{eq:29}
R_\mathrm{FF}=\frac{4\lambda G_o} {\pi^2} \sqrt{\frac{\alpha_\mathrm{E}}{1-\gamma_\mathrm{A}}},
\end{equation}
where $\lambda$ is the wavelength, $G_o$ is the antenna gain, $\alpha_\mathrm{E}=0.06$ is a fitting coefficient that is the same for all antennas and $\gamma_\mathrm{A}$ is the antenna gain reduction factor defined by the user. The starting point of the corresponding far-field region for a required error magnitude of the antenna gain defined by $1-\gamma_\mathrm{A}$.

\section{Results}
\subsection{Specific assumptions}
For the results in this section, we specialize our scatterers to identical dipoles. The scatterers are assumed to be identical vertically polarized resonant dipoles. The spatially averaged scattering cross-section of dipoles is given as~\cite{peebles1984bistatic}:
\begin{equation}
\label{eq:7}
\frac{\langle \sigma_\mathrm{dip}\rangle}{\lambda^2}=\frac{1.178\frac{L}{\lambda}+0.179\ln(22.368\frac{L}{\lambda})-0.131}{\ln^2(22.368\frac{L}{\lambda})},
\end{equation}
where $L=n\lambda/2$, $n=\{1,2,...\}$, and $\lambda$ is the free-space wavelength. Values of the averaged cross-section for different lengths of dipole that are used in the results in this section are summarized in Table~\ref{tab:sigmadipole}.

\begin{table}[t]
	\centering
	\caption{Spatially averaged scattering cross-section of dipole of length $L$}
	\label{tab:sigmadipole}
	\begin{tabular}{|l|lllll|}
		\hline
		$L/\lambda$                           & $0.5$    & $1.5$    & $2.5$    & $3.5$    & $4.5$ \\ \hline
		$\langle\sigma_\mathrm{dip}/\lambda^2\rangle$ & $0.1527$ & $0.1835$ & $0.2183$ & $0.2510$ & $0.2819$\\ \hline
	\end{tabular}
\end{table}

\begin{figure}[!t]
	\centering
	\includegraphics[width=0.9\columnwidth]{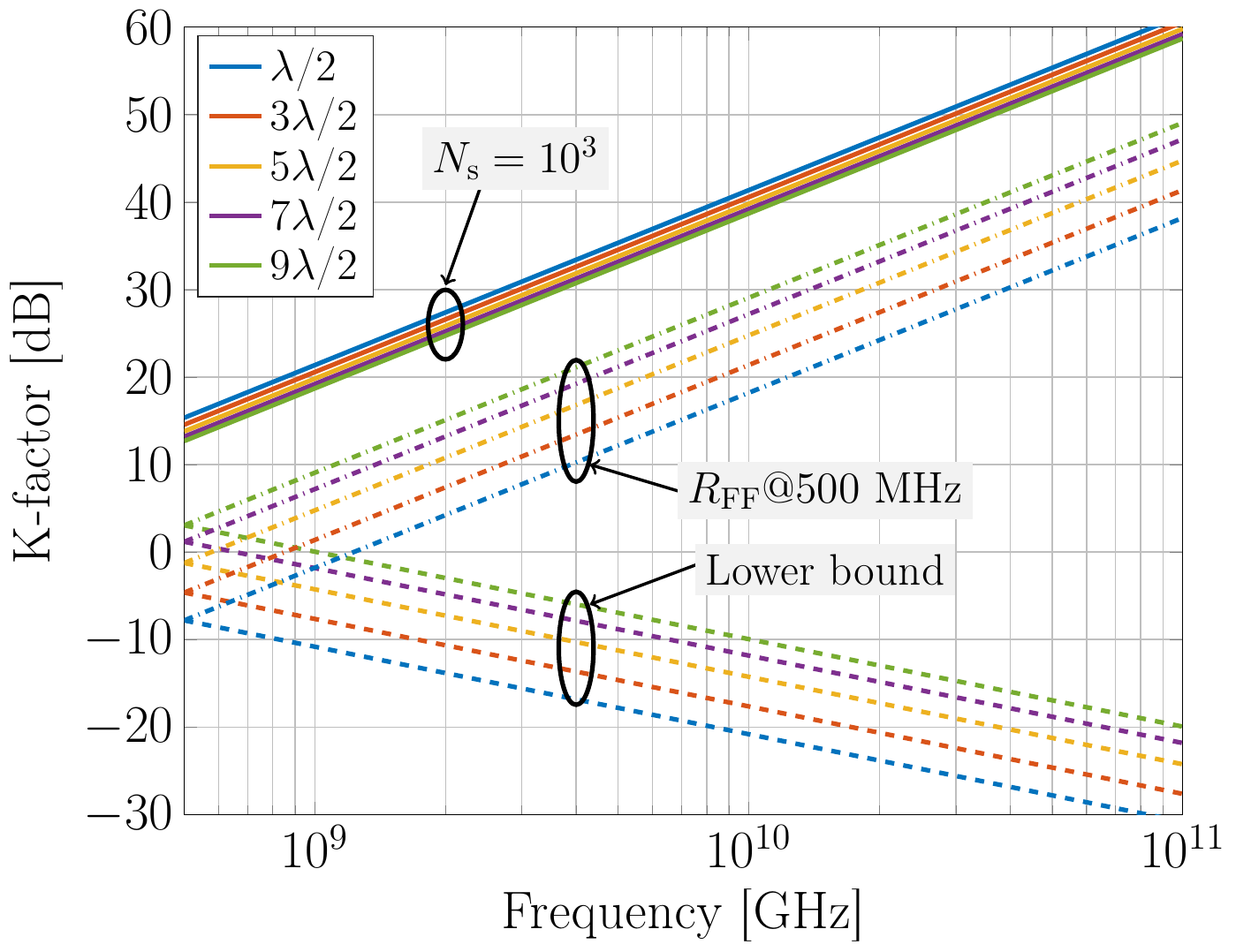}
	\caption{\small The Rician $K$-factor vs. frequency for different electrical lengths of the scatterers in a volume with $R_\text{S}=15$~m. Three different scenarios: (1) The lower bound, where the scattering volume is filled with largest possible number of scatterers where all of them are in the far-field region of each other, (2) the spacing between the scatterers is chosen based on the far-field distance ($R_\text{FF}$) at the lowest frequency, and (3) a fixed number of scatterers randomly distributed in the scattering environment, regardless of the electrical size.}
	\label{fig:cases}
\end{figure}

Assuming $R_\text{S}=15$~m, the lower bound of the $K$-factor according to \eqref{eq:28} is plotted in Fig.~\ref{fig:cases} vs. frequency, for different electrical lengths of the scatterers. It can be observed that with the increase in frequency, a larger number of scatterers can fit in the volume [according to \eqref{eq:29}] which results in the increase of the scattered power and the decrease of the $K$-factor. On the other hand, the larger electrical size of the scatterers leads to increased $K$-factor, since it means that a smaller number of scatterers can fit in the volume. However, if the number of scatterers is kept constant, the $K$-factor will increase with frequency. This is shown in Fig.~\ref{fig:cases} with the dash-dotted curves, where the spacing between the scatterers is chosen according to the $R_\text{FF}$ value at the lowest frequency, i.e., $500$~MHz. Note that in this case $N_\text{s}$ is dependent on the electrical size of the scatterers. We can also assume a case where a fixed $N_\text{s}$ is chosen for all values of the electrical length. This case is shown in Fig.~\ref{fig:cases} with the solid lines for  $N_\text{s}=1000$. Unlike the previous cases, as expected, we observe in this case that larger electrical size will lead to lower $K$-factor.

\subsection{Comparison with numerical MOM-based simulations}
We have used MoM in order to numerically simulate the model. Effects of multiple scattering and mutual coupling between the scatterers are included in the full-wave simulations. The scatterers are modeled as half-wave PEC strips of $\lambda/100$ width, which are uniformly distributed in a cubic volume with side length of $30$~m. Assuming the transmitter antenna is located far from the scatterers and the receiver antenna, it is modeled as a plane wave impinging on the volume. Furthermore, in order to remove the effect of the receiver antenna's radiation pattern, it is assumed to be an ideal omnidirectional vertically polarized antenna. This implies that in the simulations, the vertical component of the field is studied and in the model we have $D_\text{or}=1$. Finally, in order to reduce the computation time, Characteristic Basis Functions Method (CBFM) is employed for the resonant scatterers \cite{maaskant2010analysis}.

\begin{figure}[!t]
	\centering
	\includegraphics[width=0.887\columnwidth]{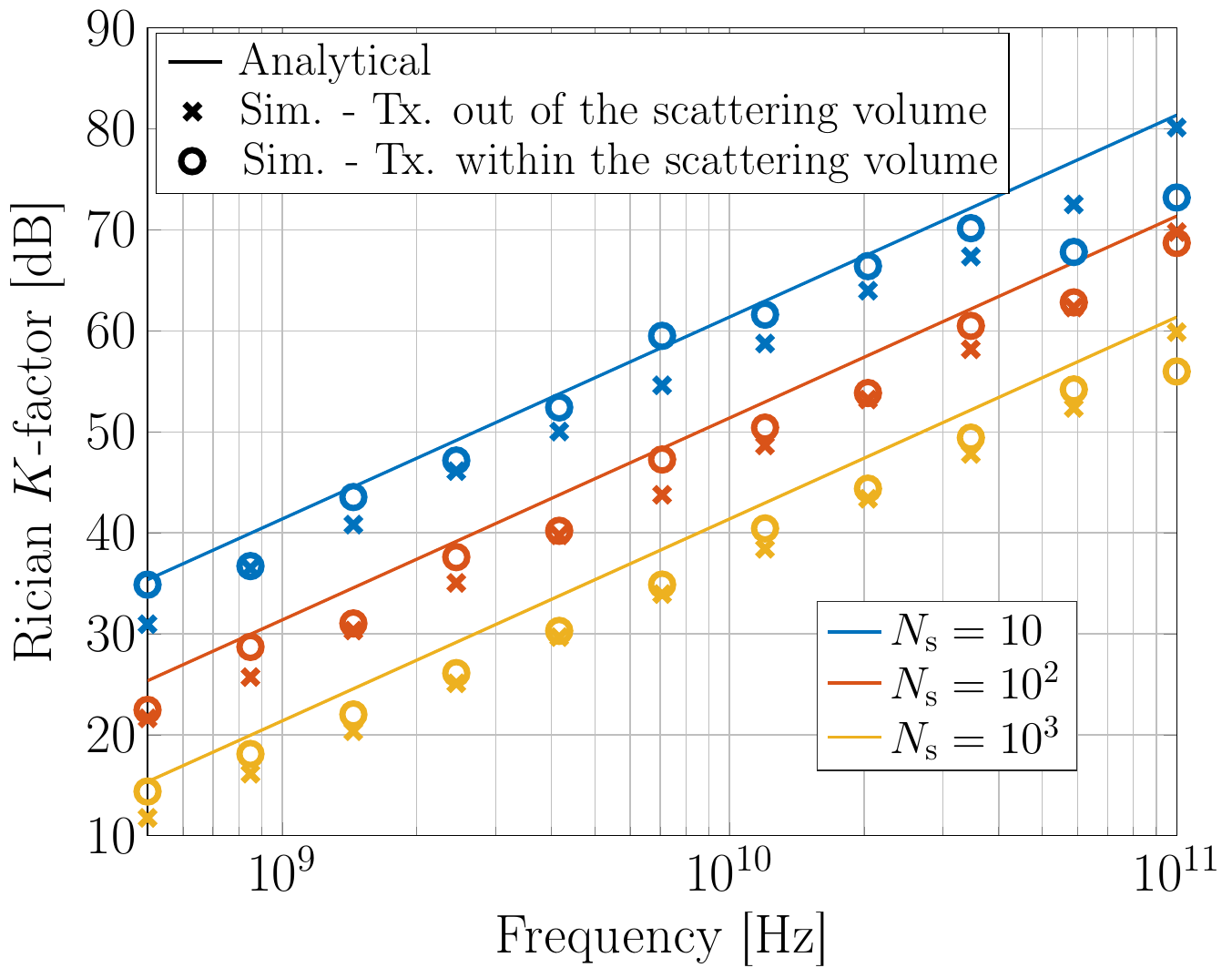}
	\caption{\small Comparison of analytical and simulated $K$-factor vs. frequency, for different numbers of identical vertical resonant scatterers. Simulated $K$-factor for the case of transmitter antenna located in the scattering volume is plotted with dashed line.}
	\label{fig:Comparison_MoM}
\end{figure}

Fig.~\ref{fig:Comparison_MoM} shows the analytical and simulated $K$-factor for cases of $10$, $100$, and $1000$ scatterers in the frequency range from $500$~MHz up to $100$~GHz. It is observed that the simulations and analytical formulas follow the same trend. However, the $K$-factor in the analytical model is slightly larger than the simulation results. This can be explained by the fact that the {\color{black}analytical} model assumes only single scattering, whereas the effect of multiple scattering is included in the simulation results. Another source of error can be the fact that the density of the scatterers is lower in the cubic volume than the spherical one. We observe that for a fixed number of scatterers, the $K$-factor increases with frequency, meaning that the LOS component becomes more dominant at the higher frequencies.

So far, we have made the assumption that the transmitter antenna is further away from the scattering volume. However, we can also assume it to be located in the scattering volume among the scatterers. We simulate the transmitter as a half wave dipole located at $R_\text{t}=R_\text{s}/2=7.5$~m. The other details of the simulation are kept as before. The simulated $K$-factor for this scenario is also plotted in Fig.~\ref{fig:Comparison_MoM}. We observe that while the $K$-factor has slightly increased, the trend of its frequency dependence does not change. Table~\ref{tab:MSE} summarizes the root-mean-square deviation of the $K$-factor values in dB, between the analytical model and the two cases in MoM simulations.

\begin{table}[t]
	\centering
	\caption{Root-mean-square deviation of the $K$-factor in dB, for the analytical model compared to MoM results}
	\label{tab:MSE}
	\begin{tabular}{|r|ccc|}
		\hline
		$N$    & $10$ & $100$ & $1000$ \\ \hline
		Tx. out of the scattering volume & $3.76$ & $4.04$ & $4.01$\\
		Tx. within the scattering volume & $3.96$ & $2.77$  & $3.08$ \\ \hline
	\end{tabular}
\end{table}

\begin{figure}[!t]
	\centering
	\includegraphics[width=0.9\columnwidth]{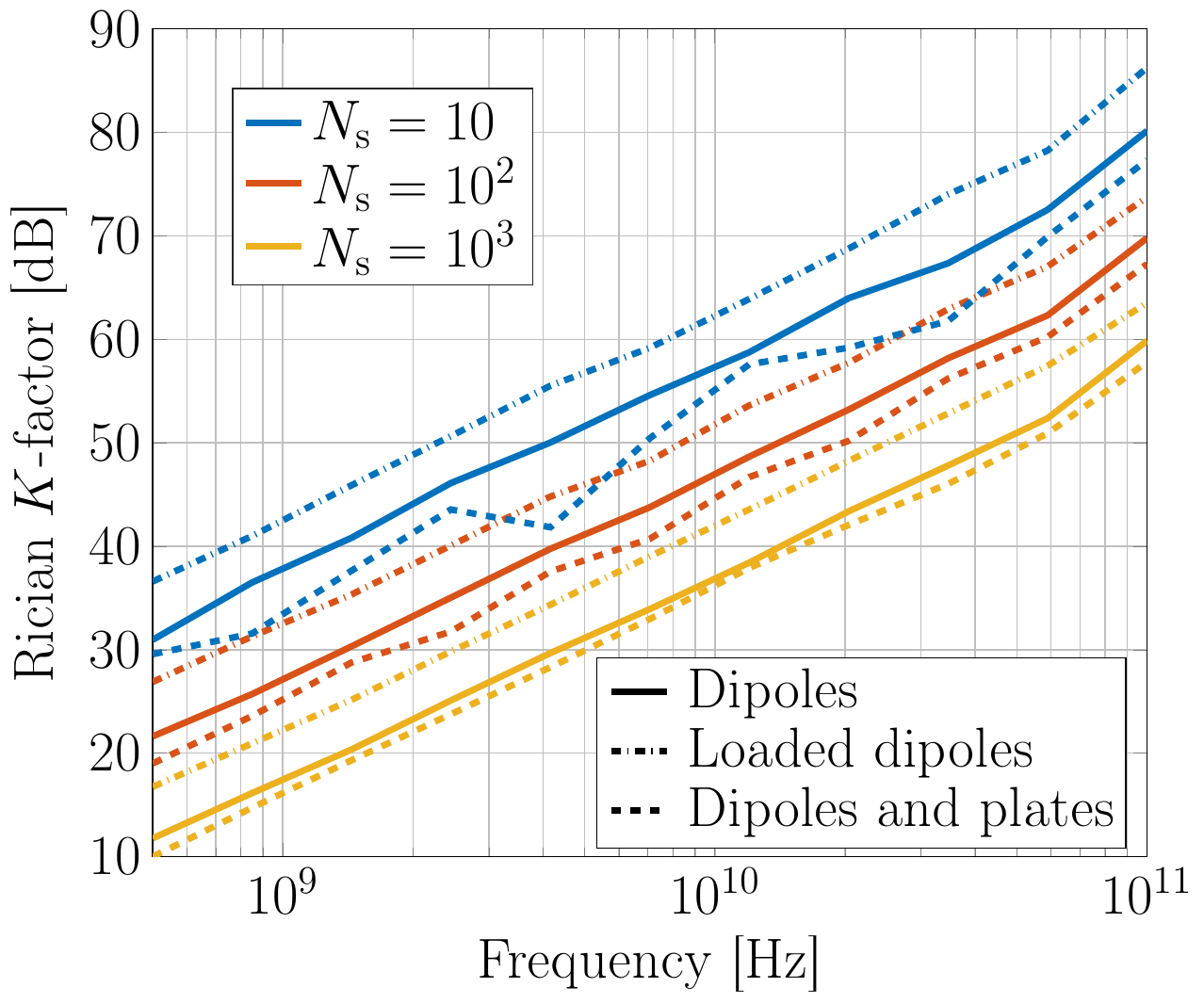}
	\caption{\small The simulated Rician $K$-factor in the presence of PEC plate scatterers and match-loaded half-wave dipoles.}
	\label{fig:PlatesLoad}
\end{figure}

In addition to the case of half-wave PEC scatterers, we have considered two other cases in the MoM simulations for the sake of comparison. In one case, scatterers are loaded with a matched load in order to absorb part of the radiated field from the transmitter antenna and to reduce the scattered power. In the second case, a number of $2\lambda\times 2\lambda$ PEC plates are also distributed among the half-wave dipoles, in order to increase the scattered power. In this case, the number of PEC plates is 1, 5, and 20 in the cases where $N_\text{S}$ is 10, 100, and 1000, respectively. 63 characteristic basis functions are used for each plate.
The $K$ factor is plotted vs. frequency for these three cases in Fig.~\ref{fig:PlatesLoad}. As expected, we observe in this figure that for the same $N_\text{S}$, terminating the dipole in matched load will lead to higher $K$-factor, while the presence of the PEC plates leads to decreased $K$-factor. However, we observe that the general trend of the $K$-factor vs. frequency is the same for all cases.

\section{Conclusions}
A simple model is introduced to investigate the frequency dependence of Rician $K$-factor in generic random scattering environments. The $K$-factor is derived analytically as a function of the average scattering cross-section of the scatterers. The formulas are verified against full-wave MoM simulations which shows a good agreement between the two. The main contributing factors to the $K$-factor are shown to be the density of the scatterers $\rho_\text{s}$, the radius of the scattering environment $R_\text{s}$, the average bistatic cross section of the scatterers $\langle \sigma_\mathrm{s}\rangle$, and the directivity of the receiving antenna $D_\mathrm{or}$. Of these, $\langle \sigma_\mathrm{s}\rangle$ and $D_\mathrm{or}$ are frequency-dependent and contribute to the frequency dependence of the $K$-factor. In the simulations two scenarios are investigated with the transmitter antenna both within the scattering volume and out of it. It is observed that for thin wire scatterers and planar scatterers, the $K$-factor increases quadratically with the frequency. Although we have used resonant dipoles in the model, the formulas have the flexibility to accommodate other types of scatterers as long as the average cross-section is known.

{\color{black}As a final remark we may say that the model presented here needs further study to take into account other specific propagation scenarios of great relevance known as obstructed line-of-sight, and the more general ones described as non-line-of-sight scenarios. Specific applications such as vehicle-to vehicle, or massive multiple-input multiple-output systems need to be studied too.}

\section*{Appendix A}
Although the scatterers are assumed to have same orientation, and due to their random positions, the polarization vector $\boldsymbol{\hat{p}}_{\mathrm{s}}$ of the scattered field impinging on the receive antenna will be randomly mismatched to its polarization vector $\boldsymbol{\hat{p}}_{\mathrm{r}}$, where $\angle\boldsymbol{\hat{p}}_{\mathrm{r}}\boldsymbol{\hat{p}}_{\mathrm{s}}$ is the mismatch angle. We have
\begin{equation}
\begin{aligned}
\Big\langle \frac{2\lambda}{j\eta I} \boldsymbol{G}_{\mathrm{r}}\!\cdot\!\mathcal{S}_n\!\cdot\!\boldsymbol{G}_{\mathrm{t}}\Big\rangle&=\Big\langle \frac{2\lambda}{j\eta I} |\boldsymbol{G}_{\mathrm{r}}||\mathcal{S}_n\!\cdot\!\boldsymbol{G}_{\mathrm{t}}|\cos(\angle\boldsymbol{\hat{p}}_{\mathrm{r}}\boldsymbol{\hat{p}}_{\mathrm{s}})\Big\rangle\\
&=\Big\langle \frac{2\lambda}{j\eta I} |\boldsymbol{G}_{\mathrm{r}}||\mathcal{S}_n\!\cdot\!\boldsymbol{G}_{\mathrm{t}}|\Big\rangle \Big\langle \cos(\angle\boldsymbol{\hat{p}}_{\mathrm{r}}\boldsymbol{\hat{p}}_{\mathrm{s}})\Big\rangle\\
&=0,\label{eq:a1}
\end{aligned}
\end{equation}
since $\langle \cos(\angle\boldsymbol{\hat{p}}_{\mathrm{r}}\boldsymbol{\hat{p}}_{\mathrm{s}})\rangle=0$ for $\angle\boldsymbol{\hat{p}}_{\mathrm{r}}\boldsymbol{\hat{p}}_{\mathrm{s}}$ uniformly distributed between $0$ and $2\pi$.

In order to show that the second ensemble average is also zero we now make further assumptions that simplify our computations
\begin{subequations}
\begin{align}
r_{\mathrm{t}n}&\approx r_\mathrm{o},\label{eq:a4}\\
r_{\mathrm{r}n}&=\rho_{n}, \label{eq:a5}
\end{align}
\end{subequations}
where \eqref{eq:a4} states that the transmit antenna is at a much larger distance from both the receive antenna and the scatterer as compared to the radius $R_\mathrm{s}$ that delimits the volume containing the scatterers. \eqref{eq:a5} is just a substitution of variables. Hence, we can now write
\begin{equation}
\Big\langle\frac{\mathrm{e}^{-jk(r_{\mathrm{t}n}+r_{\mathrm{r}n})}}{r_{\mathrm{r}n}r_{\mathrm{t}n}}\Big\rangle\! \approx\!\frac{1}{r_{o}}\Big\langle\frac{\mathrm{e}^{-jkr_{\mathrm{t}n}}\mathrm{e}^{-jk\rho_n}}{\rho_n}\Big\rangle.
\label{eq:a6}
\end{equation}
Without losing generality, we assume the local coordinate system on the receiver antenna is chosen such that the transmitter antenna is located at $\theta=0$ and the $n$-th scatterer is located at $\theta_n$. Then, \eqref{eq:a6} is approximated as
\begin{equation}
\begin{aligned}
\frac{1}{r_{o}}\Big\langle\frac{\mathrm{e}^{-jkr_{\mathrm{t}n}}\mathrm{e}^{-jk\rho_n}}{\rho_n}\Big\rangle\!&\approx\!
\frac{1}{r_{o}}\Big\langle\frac{\mathrm{e}^{-jk(r_{o}-\rho_n\cos\theta_n)}\mathrm{e}^{-jk\rho_n}}{\rho_n}\Big\rangle\\
&=\frac{\mathrm{e}^{-jk(r_o)}}{r_{o}}\Big\langle\frac{\mathrm{e}^{-jk\rho_n(1-\cos\theta_n)}}{\rho_n}\Big\rangle,
\end{aligned}
\label{eq:a6n}
\end{equation}
where, using \eqref{eq:6}, it can be shown after straightforward algebraic manipulations that
\begin{equation}
\begin{aligned}\label{eq:a7}
\Big\langle\frac{\mathrm{e}^{-jk\rho_n(1-\cos\theta_n)}}{\rho_n}\Big\rangle\!&=\!\int_{\rho=0}^{R_\mathrm{s}}\int_{\theta=0}^{\pi}\frac{\mathrm{e}^{-jk\rho(1-\cos\theta)}}{\rho}
\frac{3\rho^2}{R_\mathrm{s}^3}\frac{\sin\theta}{2}\mathrm{d}\theta\mathrm{d}\rho
\\&=0+\mathcal{O}( R_\mathrm{s}^{-2}) \text{ as }R_\mathrm{s}\rightarrow \infty.
\end{aligned}
\end{equation}

In practice, it suffices that $R_\mathrm{s}\gg\frac{\lambda}{2\pi}$.

\section*{Appendix B}
Let's first recall some antenna parameters following the definitions in \cite{kildal2015foundations}.
The directivity of an antenna is defined as
\begin{align}
D_\mathrm{o}(\boldsymbol{\hat{r}})&=\frac{4\pi|\boldsymbol{G}(\boldsymbol{\hat{r}})|^2}{2\eta P_\mathrm{rad}},\label{eq:b1}
\end{align}
where $\boldsymbol{G}(\boldsymbol{\hat{r}})$ is the far-field function satisfying the normalization integral
\begin{align}
\oint D_\mathrm{o}(\boldsymbol{\hat{r}})\mathrm{d}\Omega&=\oint\frac{4\pi|\boldsymbol{G}(\boldsymbol{\hat{r}})|^2}{2\eta P_\mathrm{rad}}\mathrm{d}\Omega=4\pi.\label{eq:b2}
\end{align}
The total radiated power is
\begin{align}
P_\mathrm{rad}&=e_\mathrm{rad}P_\mathrm{t},\label{eq:b3}
\end{align}
where $e_\mathrm{rad}$ is the radiation efficiency of the antenna and $P_\mathrm{t}$ is the input power to the antenna, which in turn is related to the current at the antenna input port $|I|$ (in \eqref{eq:13}) as
\begin{align}
P_\mathrm{t}&=\frac{1}{2}R_{\mathrm{ar}}|I|^2,\label{eq:b4}
\end{align}
The antenna gain is given by
\begin{align}
G_\mathrm{o}&=e_\mathrm{rad}D_\mathrm{o}.\label{eq:b5}
\end{align}

Following the assumption that the polarization of the scattered field at the position of the receiver antenna is random, we write the ensemble average of $|V_{\mathrm{oc}}^{\mathrm{s}}|^2$ as
\begin{equation}
\begin{aligned}
\langle|V_{\mathrm{oc}}^{\mathrm{s}}|^2\rangle\!&=\!\sum_{n=1}^{N_\mathrm{s}} \sum_{n^\prime=1}^{N_\mathrm{s}}\Big\langle \left(\!\frac{2\lambda}{\eta |I|}\!\right)^2 \left(\boldsymbol{G}_{\mathrm{r}}\!\cdot\!\mathcal{S}_n\!\cdot\!\boldsymbol{G}_{
\mathrm{t}}\right)\left(\boldsymbol{G}_{\mathrm{r}}\!\cdot\!\mathcal{S}_{n^\prime}\!\cdot\!\boldsymbol{G}_{
\mathrm{t}}\right)^\ast\Big\rangle\\
& \quad\times \Big\langle\frac{\mathrm{e}^{-jk( r_{\mathrm{r}n}+r_{\mathrm{t}n}-r_{\mathrm{r}n^{\prime}}-r_{\mathrm{t}n^{\prime}}) }}{r_{\mathrm{r}n}r_{\mathrm{t}n}r_{\mathrm{r}n^{\prime}}r_{\mathrm{t}n^{\prime}}}\Big\rangle\\
&=\!\sum_{n=1}^{N_\mathrm{s}} \Big\langle \left(\frac{2\lambda}{r_{o}\eta |I|}\right)^2 |\boldsymbol{G}_{\mathrm{r}}\!\cdot\!\mathcal{S}_n\!\cdot\!\boldsymbol{G}_{
\mathrm{t}}|^2\Big\rangle \Big\langle\frac{1}{{\rho_n}^2}\Big\rangle, \label{eq:b7}
\end{aligned}
\end{equation}
where we have used the results in Appendix A, i.e., for $n\neq n^{\prime}$ the ensemble averages are null. All the terms in \eqref{eq:b7} are identical since the scatterers are identical.

Let's denote the first ensemble average in \eqref{eq:b7} by $X$, then
\begin{equation}
\begin{aligned}
X&=\Big\langle \left(\frac{2\lambda}{r_{o}\eta |I|}\right)^2 |\boldsymbol{G}_{\mathrm{r}}\!\cdot\!\mathcal{S}_n\!\cdot\!\boldsymbol{G}_{\mathrm{t}}|^2\Big\rangle\\
&=\Big\langle \left(\frac{2\lambda}{r_{o}\eta |I|}\right)^2 |\boldsymbol{G}_{\mathrm{r}}|^2|\mathcal{S}_n\!\cdot\!\boldsymbol{G}_{\mathrm{t}}|^2 \cos^2(\angle\boldsymbol{\hat{p}}_{\mathrm{r}}\boldsymbol{\hat{p}}_{\mathrm{s}})\Big\rangle\\
&=\Big\langle \left(\frac{2\lambda}{r_{o}\eta |I|}\right)^2 |\boldsymbol{G}_{\mathrm{r}}|^2\Big\rangle \Big\langle|\mathcal{S}_n\!\cdot\!\boldsymbol{G}_{\mathrm{t}}|^2\rangle \langle\cos^2(\angle\boldsymbol{\hat{p}}_{\mathrm{r}}\boldsymbol{\hat{p}}_{\mathrm{s}})\Big\rangle, \label{eq:b10}
\end{aligned}
\end{equation}
where the factorization of the ensemble average has been performed by grouping terms that are statistically independent.

The bistatic scattering cross section is defined as
\begin{align}
\sigma(-\boldsymbol{\hat{r}}_{
\mathrm{r}n},\boldsymbol{\hat{r}}_{\mathrm{t}n})&=4\pi\frac{|\mathcal{S}_n(-\boldsymbol{\hat{r}}_{\mathrm{r}n},\boldsymbol{\hat{r}}_{\mathrm{t}n})\!\cdot\!\boldsymbol{G}_{\mathrm{t}}|^2}{|\boldsymbol{G}_{\mathrm{t}}|^2}. \label{eq:b11}
\end{align}

Then, inserting \eqref{eq:b11} into \eqref{eq:b10} gives
\begin{align}
X&=\Big\langle \left(\frac{2\lambda}{r_{o}\eta |I|}\right)^2 |\boldsymbol{G}_{\mathrm{r}}|^2\Big\rangle \frac{\langle \sigma \rangle}{4\pi} \Big\langle|\boldsymbol{G}_{\mathrm{t}}|^2\Big\rangle \Big\langle\cos^2(\angle\boldsymbol{\hat{p}}_{\mathrm{r}}\boldsymbol{\hat{p}}_{\mathrm{s}})\Big\rangle. \label{eq:b12}
\end{align}
We now compute the ensemble average terms in \eqref{eq:b10}. Since we assume that $\angle\boldsymbol{\hat{p}}_{\mathrm{r}}\boldsymbol{\hat{p}}_{\mathrm{s}}$ is uniformly distributed between $0$ and $2\pi$, then
\begin{align}
\langle\cos^2(\angle\boldsymbol{\hat{p}}_{\mathrm{r}}\boldsymbol{\hat{p}}_{\mathrm{s}})\rangle=\frac{1}{2}. \label{eq:b13}
\end{align}
Given assumption \eqref{eq:a4}, we use \eqref{eq:b1}, \eqref{eq:b3} and \eqref{eq:b5} to arrive at
\begin{align}
\langle|\boldsymbol{G}_{\mathrm{t}}|^2\rangle&\approx|\boldsymbol{G}_{\mathrm{t}}|^2=\frac{2\eta G_{\mathrm{ot}}P_\mathrm{t}}{4\pi}. \label{eq:b14}
\end{align}
Combining \eqref{eq:b2}, \eqref{eq:b3} and \eqref{eq:b4} into the first ensemble average in \eqref{eq:b12} we get
\begin{subequations}
\begin{align}
\Big\langle \left(\frac{2\lambda}{r_{o}\eta|I|}\right)^2 |\boldsymbol{G}_{\mathrm{r}}|^2\Big\rangle&=
\Big\langle \left(\frac{2\lambda}{r_{o}\eta }\right)^2 \frac{e_\mathrm{rad}R_{\mathrm{ar}}|\boldsymbol{G}_{\mathrm{r}}|^2}{2P_{\mathrm{rad}}}\Big\rangle, \label{eq:b15}\\
&=\frac{8R_{\mathrm{ar}}e_\mathrm{rad}}{2\eta}\left(\frac{\lambda}{r_{o}}\right)^2 \Big\langle \frac{|\boldsymbol{G}_{\mathrm{r}}|^2}{2\eta P_{\mathrm{rad}}}\Big\rangle, \label{eq:b16}\\
&=\frac{8R_{\mathrm{ar}}e_\mathrm{rad}}{2\eta}\left(\frac{\lambda}{r_{o}}\right)^2 \oint \frac{|\boldsymbol{G}_{\mathrm{r}}|^2}{2\eta P_{\mathrm{rad}}}\frac{\mathrm{d}\Omega}{4\pi}, \label{eq:b17}\\
&=\frac{8R_{\mathrm{ar}}e_\mathrm{rad}}{4\pi2\eta}\left(\frac{\lambda}{r_{o}}\right)^2 , \label{eq:b18}
\end{align}
\end{subequations}
where we have used the assumption that the AoA of the scattered waves are isotropically distributed since the scatterers are uniformly distributed within the spherical scattering volume.

For the second ensemble average in \eqref{eq:b7} and by using \eqref{eq:6} we straightforwardly obtain that
\begin{align}
\Big\langle\frac{1}{\rho_n^2}\Big\rangle\!&=\!\int_{0}^{R_\mathrm{s}}\frac{1}{\rho^2}\frac{3\rho^2}{R_\mathrm{s}^3}\mathrm{d}\rho=\frac{3}{R_\mathrm{s}^2}. \label{eq:b19}
\end{align}
Combining \eqref{eq:b12}, \eqref{eq:b13}, \eqref{eq:b14}, \eqref{eq:b18}, \eqref{eq:b19} into \eqref{eq:b7} provides the result in \eqref{eq:23}.

\bibliographystyle{IEEEtran}

\begin{thebibliography}{10}
\bibitem{boccardi2014five}
F.~Boccardi, R.~W. Heath, A.~Lozano, T.~L. Marzetta, and P.~Popovski, ``Five
  disruptive technology directions for {5G},'' \emph{IEEE Communications
  Magazine}, vol.~52, no.~2, pp. 74--80, 2014.

\bibitem{larsson2014massive}
E.~G. Larsson, O.~Edfors, F.~Tufvesson, and T.~L. Marzetta, ``Massive {MIMO}
  for next generation wireless systems,'' \emph{IEEE Communications Magazine},
  vol.~52, no.~2, pp. 186--195, 2014.

\bibitem{marzetta2010noncooperative}
T.~L. Marzetta, ``Noncooperative cellular wireless with unlimited numbers of
  base station antennas,'' \emph{IEEE Transactions on Wireless Communications},
  vol.~9, no.~11, pp. 3590--3600, 2010.

\bibitem{rusek2013scaling}
F.~Rusek, D.~Persson, B.~K. Lau, E.~G. Larsson, T.~L. Marzetta, O.~Edfors, and
  F.~Tufvesson, ``Scaling up {MIMO}: Opportunities and challenges with very
  large arrays,'' \emph{IEEE Signal Processing Magazine}, vol.~30, no.~1, pp.
  40--60, 2013.

\bibitem{glazunov2012mimo}
A.~A. Glazunov, V.-M. Kolmonen, and T.~Laitinen, ``{MIMO} over-the-air
  testing,'' \emph{LTE-Advanced and Next Generation Wireless Networks: Channel
  Modelling and Propagation}, pp. 411--441, 2012.

\bibitem{ngo2014aspects}
H.~Q. Ngo, E.~G. Larsson, and T.~L. Marzetta, ``Aspects of favorable
  propagation in massive {MIMO},'' in \emph{Proceedings of the 22nd European
  Signal Processing Conference (EUSIPCO 2014)}, Lisbon, Portugal, September
  2014, pp. 76--80.

\bibitem{bjornson2015massive}
E.~Bj{\"o}rnson, E.~G. Larsson, and T.~L. Marzetta, ``Massive {MIMO}: Ten myths
  and one critical question,'' \emph{IEEE Communications Magazine}, vol.~54,
  no.~2, pp. 114--123, 2016.

\bibitem{gao2015massive}
X.~Gao, O.~Edfors, F.~Rusek, and F.~Tufvesson, ``Massive {MIMO} performance
  evaluation based on measured propagation data,'' \emph{IEEE Transactions on
  Wireless Communications}, vol.~14, no.~7, pp. 3899--3911, 2015.

\bibitem{kildal2013rethinking}
P.-S. Kildal, ``Rethinking the wireless channel for {OTA} testing and network
  optimization by including user statistics: {RIMP}, pure-{LOS}, throughput and
  detection probability,'' in \emph{Proceedings of the 2013 International
  Symposium on Antennas and Propagation (ISAP)}, Nanjing, China, October 2013.

\bibitem{kildal2013new}
P.~Kildal and J.~Carlsson, ``New approach to {OTA} testing: {RIMP} and
  pure-{LOS} reference environments \& a hypothesis,'' in \emph{Proceedings of
  the 7th European Conference on Antennas and Propagation (EuCAP 2013)},
  Gothenburg, Sweden, April 2013, pp. 315--318.

\bibitem{maaskant2010analysis}
R.~Maaskant, ``Analysis of large antenna systems,'' Ph.D. dissertation,
  Technische Universiteit Eindhoven, 2010.

\bibitem{5g-mmmagic}
K.~Haneda, S.~L.~H. Nguyen, A.~Karttunen, J.~J{\"a}rvel{\"a}inen, A.~Bamba,
  R.~D'Errico, J.~Medbo, F.~Undi, S.~Jaeckel, N.~Iqbal, J.~Luo, M.~Rybakowski,
  C.~Diakhate, J.-M. Conrat, A.~Naehring, S.~Wu, A.~Goulianos, and E.~Mellios,
  ``Measurement results and final channel models for preferred suitable
  frequency ranges,'' Tech. Rep. H2020-ICT-671650-mmMAGIC/D2.2, May 2017.
  [Online]. Available: \url{http://5g-mmmagic.eu/}

\bibitem{jakes1974microwave}
W.~Jakes, \emph{Microwave mobile communications}, ser. IEEE Press classic
  reissue.\hskip 1em plus 0.5em minus 0.4em\relax IEEE Press, 1974.

\bibitem{molisch2012wireless}
A.~Molisch, \emph{Wireless Communications}, ser. Wiley - IEEE.\hskip 1em plus
  0.5em minus 0.4em\relax Wiley, 2012.

\bibitem{tsang2004scattering}
L.~Tsang, J.~Kong, and K.~Ding, \emph{Scattering of Electromagnetic Waves,
  Theories and Applications}, ser. Scattering of Electromagnetic Waves.\hskip
  1em plus 0.5em minus 0.4em\relax Wiley, 2004.

\bibitem{bialkowski2006investigating}
M.~E. Bialkowski, P.~Uthansakul, K.~Bialkowski, and S.~Durrani, ``Investigating
  the performance of {MIMO} systems from an electromagnetic perspective,''
  \emph{Microwave and Optical Technology Letters}, vol.~48, no.~7, pp.
  1233--1238, 2006.

\bibitem{kim2013generalised}
I.~Kim, S.~Xu, and Y.~Rahmat-Samii, ``Generalised correction to the friis
  formula: quick determination of the coupling in the {Fresnel} region,''
  \emph{IET Microwaves, Antennas \& Propagation}, vol.~7, no.~13, pp.
  1092--1101, 2013.

\bibitem{Riaz2015}
M.~Riaz, N.~M. Khan, and S.~J. Nawaz, ``A generalized {3-D} scattering channel
  model for spatiotemporal statistics in mobile-to-mobile communication
  environment,'' \emph{IEEE Transactions on Vehicular Technology}, vol.~64,
  no.~10, pp. 4399--4410, Oct 2015.

\bibitem{Ghoraishi2006}
M.~Ghoraishi, J.~Takada, and T.~Imai, ``Identification of scattering objects in
  microcell urban mobile propagation channel,'' \emph{IEEE Transactions on
  Antennas and Propagation}, vol.~54, no.~11, pp. 3473--3480, Nov 2006.

\bibitem{Samimi2016}
M.~K. Samimi and T.~S. Rappaport, ``{3-D} millimeter-wave statistical channel
  model for {5G} wireless system design,'' \emph{IEEE Transactions on Microwave
  Theory and Techniques}, vol.~64, no.~7, pp. 2207--2225, July 2016.

\bibitem{van2007electromagnetic}
J.~Van~Bladel, \emph{Electromagnetic Fields}, ser. IEEE Press Series on
  Electromagnetic Wave Theory.\hskip 1em plus 0.5em minus 0.4em\relax John
  Wiley and Sons, 2007.

\bibitem{kildal2015foundations}
P.-S. Kildal, \emph{{Foundations of Antennas: A Unified Approach} for
  {Line-of-Sight} and {Multipath}}.\hskip 1em plus 0.5em minus 0.4em\relax
  Kildal Antenn AB, 2015, \textit{Available at www.kildal.se.}

\bibitem{jaeger1992physics}
H.~M. Jaeger and S.~R. Nagel, ``Physics of the granular state,''
  \emph{Science}, vol. 255, no. 5051, p. 1523, 1992.

\bibitem{peebles1984bistatic}
P.~Z. Peebles, ``Bistatic radar cross sections of chaff,'' \emph{IEEE
  Transactions on Aerospace and Electronic Systems}, no.~2, pp. 128--140, 1984.

\end{thebibliography}

\end{document}